\documentclass[14pt]{article}
\pdfoutput=1
\usepackage{amsmath,amssymb,amsfonts,amsthm,bm,bbm,cancel,wasysym}
\usepackage{epsfig,graphics,graphicx,epstopdf}
\usepackage{array,booktabs,colortbl,colordvi,multirow}
\usepackage{colordvi,color,xcolor}
\usepackage{hyperref}
\usepackage{rotating}
\usepackage{appendix}

\usepackage{verbatim}
\usepackage{cite}
\usepackage{subfig}
\usepackage{setspace}
\usepackage{url}
\usepackage[percent]{overpic}
\usepackage{slashed}
\usepackage{authblk}
\usepackage{xspace}
\usepackage{fullpage}
\usepackage{hyperref}

\def\ie{{\it i.e.}}
\def\eg{{\it e.g.}}
\def\etc{{\it etc}}

\def\to{\rightarrow}

\newskip\zatskip \zatskip=0pt plus0pt minus0pt
\def\matth{\mathsurround=0pt}
\def\lsim{\mathrel{\mathpalette\atversim<}}
\def\gsim{\mathrel{\mathpalette\atversim>}}
\def\atversim#1#2{\lower0.7ex\vbox{\baselineskip\zatskip\lineskip\zatskip
  \lineskiplimit 0pt\ialign{$\matth#1\hfil##\hfil$\crcr#2\crcr\sim\crcr}}}

\begin{document}

\begin{flushright}
SLAC-PUB-17678\\
\today
\end{flushright}
\vspace*{5mm}

\renewcommand{\thefootnote}{\fnsymbol{footnote}}
\setcounter{footnote}{1}

\begin{center}
{\Large {\bf Kinetic Mixing, Dark Higgs Triplets, $M_W$ and All That}}\\

\vspace*{0.75cm}

{\bf Thomas G. Rizzo}~\footnote{rizzo@slac.stanford.edu}

\vspace{0.5cm}

{SLAC National Accelerator Laboratory}\\ 
{2575 Sand Hill Rd., Menlo Park, CA, 94025 USA}

\end{center}
\vspace{.5cm}

\begin{abstract}
 
\noindent
The kinetic mixing (KM) portal is a popular mechanism which allows light dark matter (DM) in the mass range below $\sim 1$ GeV to achieve the observed relic density by thermal means and 
can be effectively described by only a few parameters, \eg, $\epsilon$, the strength of this KM. In the simplest setup, the Standard Model (SM) $U(1)_Y$ hypercharge gauge boson and a 
$U(1)_D$ dark photon (DP), which only couples to fields in the dark sector, experience KM via loops of portal matter (PM) fields which have both SM and dark charges thus generating a 
small coupling between us and dark matter (DM). However, if one wishes to understand the underlying physics behind this idea in a deeper fashion we need to take a step upward to a more 
UV-complete picture. Meanwhile, CDFII has measured a value for the $W$-boson mass which lies significantly above SM expectations. In this paper we speculate that this shift in the 
$W$'s mass may be related to the $\sim 1$ GeV mass of the DP within a framework of scalar PM that leads to phenomenologically interesting values of $\epsilon$ via non-abelian KM due 
to the existence of an $SU(2)_L$, $Y=0$, {\it complex} Higgs triplet which carries a non-zero value of the $U(1)_D$ dark charge. Possible gauge boson plus missing energy signatures of 
this scenario that can appear at the LHC and elsewhere are examined. Indeed, with modest assumptions, all of the new scalar PM states are predicted to have masses below roughly 
$\simeq 630$ GeV and so should be at least kinematically accessible. The HL-LHC will very likely to be able to explore all of this model's allowed parameter space. 
\end{abstract}

\renewcommand{\thefootnote}{\arabic{footnote}}
\setcounter{footnote}{0}
\thispagestyle{empty}
\vfill
\newpage
\setcounter{page}{1}

\newpage

\section{Introduction}

The nature of dark matter (DM) and its interactions with the Standard Model (SM) - the subject of the current paper - remains mysterious and, together with several other problems such 
as the hierarchy, the generation of neutrino masses, the possible observation of lepton non-universality, the discrepancy in the value of the muon's $g-2$ and the baryon-antibaryon 
asymmetry, shapes much of our 
current research into physics beyond the SM (BSM).  We know that the SM is incomplete - but where will we see the first real, undeniably obvious, in your face break?  We may indeed be 
surprised from which direction it actually comes - even with all of the work done in this area over many past decades. Perhaps, it is more than likely 
that to make any serious advancement on any of these problematic fronts we will need to wait until we have new input from an ever wider range of experiments and/or 
astrophysical observations to help guide us. For example, the lack of any signals so far for the traditional DM candidates, such as Weakly Interacting Massive Particles 
(WIMPS)\cite{Arcadi:2017kky,Roszkowski:2017nbc} or axions\cite{Kawasaki:2013ae, Graham:2015ouw} has led to an explosion of new ideas covering huge ranges in both DM 
masses and couplings\cite{Alexander:2016aln,Battaglieri:2017aum,Bertone:2018krk}. One path that has gained much interest is the existence of 
(renormalizable) portals wherein a messenger field connects the physics of a generalized dark sector, containing the DM and possibly other states, with that of the SM. 
Of particular interest is the kinetic mixing (KM)/vector portal scenario\cite{vectorportal,KM} where a new gauge field, the dark photon (DP)\cite{fabbrichesi2020dark}, 
which is the gauge boson associated with a new $U(1)_D$ symmetry,  
couples to only DM and other dark sector fields. This new gauge field, $V$, can experience KM with the neutral SM gauge fields via a set of vacuum polarization-like diagrams wherein 
are exchanged new scalar and/or fermion fields, here 
termed Portal Matter (PM)\cite{Rizzo:2018vlb,Rueter:2019wdf,Kim:2019oyh,Rueter:2020qhf,Wojcik:2020wgm,Rizzo:2021lob,Wojcik:2021xki,Rizzo:2022qan,Wojcik:2022rtk}, 
which carry both dark and SM quantum numbers. Being loop suppressed, the resulting strength of this interaction is rather weak and is usually described 
via a single parameter, $\epsilon \sim 10^{-4}-10^{-3}$, which, for DM and DP in the mass range $\lsim$ 1 GeV can reproduce the observed DM relic abundance\cite{Aghanim:2018eyx} 
via the conventional freeze-out mechanism while simultaneously avoiding numerous other existing experimental constraints. Generally, the DP mass itself is generated through the vev, $v_s$, 
of a dark, complex neutral SM singlet Higgs boson in an analogous fashion to what happens in the SM itself but at the $\sim 1$ GeV mass scale. Many potential new and exciting experiments 
may eventually explore the details of this range of DP/DM masses and couplings\cite{Agrawal:2021dbo}. 

While we've been hard at work and waiting, the community has recently been thrown an unexpected curve ball by the new, high-precision measurement of the $W$ boson mass, $m_W$, 
reported by the 
CDF II collaboration\cite{CDF:2022hxs}. The value that they obtained, $m_W^{CDF}=80.4335\pm 0.0094$ GeV, lies far above (by $\sim 7 \sigma$) the conventional prediction, 
$m_W^{SM}=80.3496\pm 0.0057$ GeV, obtained in the SM\cite{deBlas:2022hdk} from the updated but otherwise well-known input parameter values of 
$G_F,m_Z,\alpha_{QED}, m_t, m_{h_{SM}}$ and $\alpha_s(m_Z)$. This new result also lies significantly above the previous world average of multiple $M_W$ experimental measurements, \ie, $m_W^{old}=80.379\pm 0.012$ GeV\cite{deBlas:2022hdk}. If one carefully combines the new CDFII result with other existing data, then the new `world average' of 
$W$ mass measurements is found to be\cite{deBlas:2022hdk} $m_W^{ave}= 80.4133\pm 0.0080$ GeV, 
a result that we will make use of below, and which still lies quite far away from the SM prediction. Of course, while this new result certainly needs 
to be verified by the ATLAS and CMS collaborations at the LHC, or more likely, at future $e^+e^-$ colliders, one cannot help but to speculate what the implications of this measurement might 
be and how it could impact other areas of BSM physics. 

A traditional, but non-universal, way to express the effects of new physics sources indirectly is via the oblique parameters, $S,T,U$\cite{Peskin:1990zt,Peskin:1991sw} 
(and also via the extension including the additional obilque parameters $V,W,X$\cite{Maksymyk:1993zm}), which can adequately describe a broad class of BSM models. Subsequent 
to the new CDFII measurement, fits to just $S,T$ alone and also to all three of $S,T,U$ simultaneously have been performed\cite{deBlas:2022hdk,Lu:2022bgw,Asadi:2022xiy} 
with very interesting results. For example, it's possible that almost the entire effect may be due to a 
non-trivial value of $U\neq 0$ alone since only $m_W$ and the less precisely determined $W$ total decay width, $\Gamma_W$, probe possible non-zero values of this parameter, leaving all 
other the electroweak observables unaltered. In such a case, the difficulty is finding a model predicting a large value of $U$ but which also correspondingly keeps the predicted values of 
$S,T$ small. If $U$ is itself assumed to be small, as in most BSM models, then sizable values of $S,T$ are the natural result\cite{deBlas:2022hdk,Lu:2022bgw,Asadi:2022xiy}.  
Much theoretical effort has in particular focussed on obtaining 
a large value for the parameter $T$ which can occur either at the tree-level or via significant 1-loop radiative contributions - both of which lead to custodial symmetry breaking - with  
much of this work focused on modifications to the SM scalar sector\cite{Big} as does the analysis below. 

Historically\cite{Mathur:1977cq}, this situation was usually expressed as a violation of the well-known SM tree-level condition, $\rho=1$, where 
\begin{equation}
\rho= \frac{m_W^2}{m_Z^2c_w^2}\,,
\end{equation}
with $c_w=\cos \theta_w$, that occurs naturally in the SM with the gauge symmetries broken only by $SU(2)_L$ isodoublets plus the possible presence of Higgs isosinglets.  
$\rho=1+\delta \rho$ is then directly related to the oblique parameter $T$ as $\delta \rho=\alpha_{QED}T$ where, as noted, $\delta \rho \neq 0$ is the result of tree-level and/or large 
loop-level custodial symmetry violating effects. For example, at the tree-level there are two very well-known and well-explored ways to increase the `expected' value of $m_W$ with the 
measured value of $m_Z$ taken as a conventional input parameter: If a new $Z'$ exists, corresponding to an extended gauge symmetry, above the SM $Z$ mass but which mass mixes 
with it\cite{Hewett:1988xc,Langacker:2008yv,Leike:1998wr},  
then the SM $Z$ mass itself is `depressed' in a see-saw fashion so that, using it as an input, yields a result where $m_W$ is apparently increased compared to the naive 
SM expectations, \ie,  $\rho>1$. In such a case, the $W$'s couplings to the SM Higgs boson are unaltered at the tree-level.  
Similarly, the introduction of Higgs fields\cite{Gunion:1989we} with weak isospin $>1/2$ that have non-zero vevs can also modify the value of $\rho$, \eg, 
the introduction of a $Y=0$ scalar 
triplet with a non-zero vev, $v_t$, yields $\delta \rho =4v_t^2/v_d^2>0$, with $v_d$ being the familiar SM Higgs isodoublet vev, being a classic example\cite{Gunion:1989we} from decades 
ago\cite{Ross:1975fq}{\footnote {In such scenarios, the $W$'s coupling to the SM Higgs will receive a small alteration correlated with $\delta \rho$.}}. Naively, if we were 
to completely {\it ignore} any further corrections arising at the 1-loop level then we would find in such a case that\cite{Peskin:1990zt,Peskin:1991sw}  (here $s^2_w$ is the usual weak 
mixing angle) 
\begin{equation}
\frac{\Delta m_W^2}{m_W^2}=\frac{m^2_{W_{ave}}-m^2_{W_{SM}}}{m^2_{W_{SM}}}= \frac{4(1-s^2_w)}{1-2s^2_w}~\frac{v_t^2}{v_d^2}\,,
\end{equation}
so that using the results from above 
one finds that $v_t\simeq 4$ GeV if we employed the quoted central values of $m_W^{ave}$ and $m_W^{SM}$. Interestingly, in this `$T$-only' scenario, a shift 
in $M_W$ is directly correlated with a corresponding somewhat smallish downward shift\cite{Peskin:1990zt,Peskin:1991sw} in the value of the familiar weak mixing angle, 
$s_w^2\simeq 0.2315$, \ie, 
\begin{equation}
\frac{\Delta m_W^2}{m_W^2}=-\frac{\Delta s_w^2}{s_w^2}\,.
\end{equation}
However, it is important that we note that, in at least a simpler version of the model that we will describe below with a only single additional real isotriplet Higgs field but also having a 
similar tree-level violation of custodial symmetry, it is known that the 1-loop contributions to $\delta \rho \sim T$ can be sizable albeit, since it is now a divergent quantity and must 
be renormalized. Indeed, one finds that $T$ is now a scheme, scale and also the choice of input parameters 
dependent quantity\cite{Lynn:1990zk,Blank:1997qa,Czakon:1999ha,Forshaw:2001xq,Forshaw:2003kh,Chen:2006pb,Chankowski:2006hs,Chen:2008jg}.  
This can imply that the {\it effective} value of $v_t$ may be, \eg, somewhat lower, by perhaps up to $O(1)$ factors, than is the naive tree-level value that 
we have just obtained above; however, it remains safe to think of $v_t\sim O(1)$ GeV.  Further, it is known that when the custodial symmetry is absent, the values of the remaining usual oblique 
parameters, $S,U$ must be determined in a careful manner to avoid unphysical divergences and/or gauge-dependent results\cite{Albergaria:2021dmq}. 

Now what, if anything, has any of this to do with DM in general or, more specifically, with KM and possible light DP/DM masses near the $\sim1$ GeV scale? It is clear that a light DP 
which KM and/or mass mixes with the SM $Z$ will, if anything, push the $Z$ mass upwards (albeit by a very small amount given the tiny expected value of KM parameter $\epsilon$ and 
the small mass-squared ratio $m_V^2/M_Z^2$) so that 
$m_W$ will appear to {\it decrease} in comparison to the SM expectations (again by an unobservably small amount). This is not the effect which concerns us. Here we will explore a link 
between the DP (and hence DM) mass scale and the upward shift in the $W$ boson mass through a new, now {\it complex}, $Y=0$ triplet's vev, $v_t~\sim$ GeV.  We will argue 
that $v_t \sim v_s$, the singlet vev encountered above, and so, taking the $U(1)_D$ gauge coupling $g_D \sim e$, we find that $m_V \sim 1$ GeV is the natural outcome, \ie, the 
upward shift in SM $W$ mass is 
directly correlated with the mass of the DP. In the discussion below we will employ a rather simple toy model that has all these moving parts: a new $U(1)_D$ gauge symmetry plus an 
extended Higgs sector allowing for both $\delta \rho \sim T>0$,  
a naturally occurring $m_V~\sim 1 $ GeV related to the apparent upward shift in the $W$ mass, and a finite, calculable value for (or at least a contribution to) $\epsilon \sim O(10^{-4})$ via 
kinetic mixing without any further extension beyond the new PM scalar sector, \ie, the introduction of any additioinal fields to be employed {\it only} as PM{\footnote {Of course such 
new fields, \eg, heavy vector-like fermions with both SM and dark sector quantum numbers\cite{Rizzo:2018vlb,Rueter:2019wdf,Kim:2019oyh,Rueter:2020qhf,Wojcik:2020wgm,Rizzo:2021lob,Wojcik:2021xki,Rizzo:2022qan,Wojcik:2022rtk}, might also be 
present in a more complete version of the present setup and they would also contribute to abelian KM as usual.}}. As we will see, this model directly leads to rather unique and distinctive pattern 
of signatures for the production of these new scalar PM states and their subsequent decays at the LHC. These are found to be qualitatively similar to - but quantitatively distinct from - 
those found in the 
case on an additional pair of dark PM scalar doublets, as was discussed in our earlier work\cite{Rueter:2020qhf}, which we will generally follow and compare with in the analysis below. This is 
particularly interesting as the extended PM scalar sectors in both models have the same number of new degrees of freedom and have the same electric charge assignments. Also, in both cases, 
the masses of these new scalar PM fields are constrained from above by the familiar perturbativity and unitarity arguments.

The outline of this paper is as follows: in Section 2, we present an overview of the new PM scalar sector of our setup, presenting the various mass-squared matrices for the CP-odd, CP-even 
and charged scalar sectors which then determine the corresponding mass eigenstates and mass values, both obtained to leading order in small ratios $v_{t,s}/v_d$. Section 3 contains a discussion of a 
general scenario having both abelian and non-abelian DP KM with the SM fields. The non-abelian KM generated in the present setup by the PM/dark isotriplets after SSB and the by the inequality of the 
two physical charged scalar masses (something that is absent when the new $Y=0$ triplet field is real) is then discussed and the expected magnitude of the KM parameter, $\epsilon$, 
is then analyzed.  Section 4 contains a scan of the model parameter space as well as a survey of some of the phenomenological implications associated with the new scalar PM fields in 
the current setup for BSM searches at the LHC. A brief discussion and our conclusions are finally presented in Section 5.

\section{Model Basics: New Scalar Sector}

As noted above, the basic model building assumptions that we will make here are as follows: ($i$) At the electroweak scale and below, the SM gauge group is augmented by an additional 
$U(1)_D$ factor corresponding to the, assumed light, $\sim 1$ GeV, dark photon, $V$. The SM fields are assumed not to carry any dark charge, \ie, they have $Q_D=0$, so that before 
KM and mass mixing they do not couple to $V$.  
($ii$) The SM Higgs sector, consisting of the usual isodoublet, $\Phi$, is also augmented by a complex isosinglet, $S$, with $Q_D=1$, which traditionally solely plays the role of the dark 
Higgs through its vev, $v_s$.  This extended Higgs sector now also includes a $Y=0$, isotriplet, $\Sigma$, which {\it also} carries the same dark charge, $Q_D=1$, and whose neutral member 
obtains a vev, $v_t$, which as is very well-known causes a positive tree-level shift in the $\rho$ parameter, $\delta \rho=4v_t^2/v_d^2$, as discussed above. Note that, unlike in most standard 
treatments, $\Sigma$ is here a {\it complex} field since it carries a non-zero dark charge. This implies that its oppositely charged $T_3=\pm 1$ members are {\it no longer related} to one another 
and that the $T_3=0$ member, $\Sigma^0$, in particular, is also necessarily a 
complex field with both CP-even as well as CP-odd components. Note that the role(s) of the dark Higgs and PM are here simultaneously played by the two new scalar fields.
The triplet vev, $v_t$, as well as the singlet vev, $v_s$, which we expect to be of comparable size (as will be discussed further below), will now {\it both} contribute to 
the mass of the DP as both of these fields carry identical non-zero dark charges, $Q_D=1$. 

We can now decompose the real and imaginary parts of these three {\it complex} Higgs fields as
\begin{equation}
\Phi = \begin{pmatrix} \phi^+ \\\ \phi^0=\frac{h+v_d+ia}{\sqrt{2}} \end{pmatrix} ~~~
\Sigma = \begin{pmatrix} \Sigma_1^+ \\\ \Sigma^0=\frac{\sigma+v_t+i a_t}{\sqrt{2}} \\\ \Sigma_2^-\end{pmatrix} ~~~
S = \begin{pmatrix} \frac{s+v_s+i a_s}{\sqrt{2}} \end{pmatrix}
\end{equation}
with $v_{d,t,s}$ being the relevant vevs of the doublet, triplet and singlet Higgs fields, respectively, 
and whose interactions with themselves and each other are described by the most general scalar potential, $U$, allowed by both SM and $U(1)_D$ gauge invariance, \ie, 
\begin{equation} 
\begin{aligned}
U = & -m^2 \Phi^\dagger \Phi -m_\Sigma^2 Tr(\Sigma^\dagger \Sigma) -m_S^2 S^\dagger S +\lambda ( \Phi^\dagger \Phi)^2 +\lambda_S( S^\dagger S)^2  +\lambda_1 [Tr( \Sigma^\dagger \Sigma )]^2 + \lambda_2 Tr [(\Sigma^\dagger \Sigma )^2] \\& +\kappa \Phi^\dagger \Phi S^\dagger S + [\lambda_{4\Phi} \Phi^\dagger \Phi +\lambda_{4S} S^\dagger S]~Tr(\Sigma^\dagger \Sigma)+\lambda_5\Phi^\dagger \Sigma\Sigma^\dagger \Phi +\tilde \lambda \Phi^\dagger (\Sigma S^\dagger +\Sigma^\dagger S)\Phi\,.
\end{aligned}
\end{equation}
Here we specifically note that the presence of the $\tilde \lambda$ term, which will be of great importance below, is only possible because {\it both} $S$ and $\Sigma$ simultaneously 
carry the same dark charge, 
$Q_D=1$.  We can adjust the phases of the various terms in $U$ so that CP remains a good symmetry and thus the mass eigenstates for the CP-even and CP-odd neutral fields 
can be discussed separately. Calculations are most easily performed by employing the familiar $ 2\times 2$ representation of the isotriplet scalar field, $\Sigma$, obtained when dotted into 
the Pauli spin matrices:
\begin{equation}
\Sigma = \begin{pmatrix}  \Sigma^0/\sqrt 2 & \Sigma_1^+ \\\ \Sigma_2^- & -\Sigma^0/\sqrt 2\end{pmatrix}.
\end{equation}

As is usual, the minimization of the potential, $U$, allows us to determine the mass parameters $m^2, m_\Sigma^2$ and $m_S^2$ in terms of the various vevs and the quartic couplings 
via the following three tadpole conditions:  
\begin{equation} 
\begin{aligned}
m^2=& \lambda v_d^2+\frac{\kappa}{2} v_s^2+\frac{1}{2}(\lambda_{4\Phi}+\frac{\lambda_5}{2})v_t^2-\frac{\tilde \lambda v_tv_s}{\sqrt 2} \\
m_\Sigma^2=&\frac{1}{2}(\lambda_1+\frac{\lambda_2}{2})v_t^2+\frac{1}{2}(\lambda_{4\Phi}v_d^2+\lambda_{4S}v_s^2)+\frac{\lambda_5}{2}v_d^2-\frac{\tilde \lambda v_s v_d^2}{2\sqrt 2 v_t}\\
m_S^2=&\lambda_S v_s^2+\frac{\kappa}{2} v_d^2+\frac{1}{2} \lambda_{4S}v_t^2-\frac{\tilde \lambda v_t v_d^2}{2\sqrt 2 v_s} \,.
\end{aligned}
\end{equation}
Note the two small vevs appearing in the {\it denominators} of the last terms in the expressions for both $m_S^2$ and $m_\Sigma^2$. 
Hereafter we will employ the small vevs to SM vev ratios, $x_{t,s}=v_{t,s}/v_d\sim O(10^{-2})<<1$, so that the tree-level value of the $\rho$ parameter is now just given by $\rho=1+4x_t^2$; 
for convenience we will also define the ratio of the two small vevs themselves, $t=v_t/v_s$. 
Recall that, before mass and kinetic mixing, the DP obtains a mass via both of these dark Higgs vevs, \ie, $m^2_V=g_D^2(v_s^2+v_t^2)=g_D^2v_s^2(1+t^2)$, 
with $g_D$ being the dark gauge coupling, and so is $O(1)$ GeV for $g_D\sim e$ and $t\sim 1$. Since we want to keep the $V$ field light, one might expect that $v_s \lsim v_t$, \ie, 
$t\gsim 1$, but then 
additional constraints arise as we will discuss below. Combining the results for the tadpole conditions together with the other terms in the 
potential, $U$, as well as the real and imaginary decompositions of the fields given above allows us to fully determine the various physical scalar mass eigenstates.  It should be noted that in 
most of our discussions it will be most transparent and sufficient for our purposes to work to leading order in the two small parameters, $x_{t,s}$. 

Before progressing, it is interesting to examine what our expectations are for what the physical scalar degrees of freedom in this model will be after spontaneous symmetry breaking 
occurs. To lowest order in the $x_{t,s}$, of the neutral CP-odd fields, one of them, $a \simeq G_Z$, will become the Goldstone boson for the SM $Z$ while one a linear combination of the 
remaining CP-odd fields, 
$G_V\simeq a_s c_\phi+a_t s_\phi~${\footnote {Here, $s_\phi(c_\phi)=\sin \phi(\cos \phi)$, etc.}}, with, as we will discover, 
$t=t_\phi=s_\phi/c_\phi=v_t/v_s$, will become the Goldstone boson for the the DP field, $V$. The remaining orthogonal field combination will be realized as a heavy, physical CP-odd state, 
$A$ (not to be confused with the photon). Of the neutral CP-even states, to the same order in $x_{t,s}$, one will be identified with the $\simeq h_{SM}$ state at $\simeq 125$ GeV, a second 
with a light dark Higgs, $h_{D}$, at $\sim 1$ GeV, while the third will be another new heavy state, $H$, which we will find is essentially degenerate in mass with $A$. Similarly, of the 
charged states, to 
lowest order in the small parameters, $\phi^\pm$ will essentially become the Goldstone bosons for the $W^\pm$ as in the SM while the $\Sigma_i^\pm$ will remain as non-degenerate 
physical fields. These expectations will be realized in actuality below. 

To begin, we note that in the weak eigenstate basis for the three CP-odd fields, $a,a_s,a_t$, the mass-squared matrix generated by $U$ can be written as 
\begin{equation}
{\cal M}^2_{odd}=  \frac{\tilde \lambda v_d^2}{2\sqrt 2}     \begin{pmatrix} 0 & 0&0 \\0& v_t/v_s & -1 \\ 0& -1 & v_s/v_t\end{pmatrix},
\end{equation}
which yields the single non-zero eigenvalue associated with the lone physical CP-odd field, $A$:
\begin{equation}
M_A^2=\frac{\tilde \lambda v_d^2}{2\sqrt 2} \left( \frac{v_s}{v_t}+\frac{v_t}{v_s} \right)\,,
\end{equation}
where this mass eigenstate is indeed given by the combination $A= a_t c_\phi-a_s s_\phi$ with $s_\phi, c_\phi$ as suggested above. As noted the other two CP-odd massless fields 
are just the Goldstone bosons for the $Z$ and for the DP, $G_V$; we will frequently employ the $G_V$ notation to represent the longitudinal mode of $V$ in the Goldstone Boson 
Equivalence Theorem\cite{GBET} limit in our discussion below. 

The corresponding CP-even mass-squared matrix in the paralleling $\phi^0,s,\sigma$ weak eigenstate basis is now given by
\begin{equation}
{\cal M}^2_{even}=  \begin{pmatrix}  2\lambda v_d^2 & \kappa v_sv_d & 0 \\  \kappa v_sv_d & 2\lambda_Sv_s^2+kv_t/v_s & -k \\ 0 & -k &kv_s/v_t   \end{pmatrix},
\end{equation}
where for brevity we have here defined the commonly appearing combination, $k=\tilde \lambda v_d^2/(2\sqrt 2)$; note the absence of any direct $\phi^0-\sigma$ mixing. To lowest order in 
the two small parameters one then finds the physical masses to be given by{\footnote{Here by `$h_{SM}$' we mean the approximate SM state with a mass of $\simeq$125 GeV state 
observed at the LHC.}} 
\begin{equation} 
\begin{aligned}
m_{h_{SM}}^2 \simeq & ~2\lambda v_d^2\\
m_H^2 \simeq &~m^2_A =\frac{\tilde \lambda v_d^2}{2\sqrt 2}~\left(t+\frac{1}{t} \right )\\
m_{h_D}^2 \simeq & ~\frac{2\lambda_S v_s^2}{1+t^2}~\left(1-\frac{\kappa^2}{4\lambda \lambda_S} \right)\,,
\end{aligned}
\end{equation}
where we see that the SM Higgs mass is effectively unaltered to this order in the small parameters, $H$ and $A$ are found to be essentially degenerate at this same order and the dark 
Higgs mass, $m_{h_D}$, is found to be similar to the 
case of the simple singlet dark Higgs model except for the presence of the additional overall $(1+t^2)^{-1}$ factor with $t \neq 0$, in the present scenario.  
Recall that in the familiar singlet dark Higgs model, the parameter $\kappa$ must be kept quite small to suppress $h_{SM}-h_D$ mixing to satisfy the branching fraction 
constraint on the invisible decay of the SM Higgs, \ie, $B_{inv}\leq 0.11$\cite{ATLAS:2020kdi}, arising from the decay modes $h_{SM}\to 2h_D,2G_V$, where it is assumed that both the 
dark Higgs 
and the DP decay invisibly or to unreconstructed final states. In any case, here we need only more loosely require that $\kappa^2 <4\lambda \lambda_S$ to avoid a large mass shift that 
might drive this mass-squared term tachyonic. We will return to the issue of the invisible width of $h_{SM}$ within the present setup below.

Finally, the mass-squared matrix for the charged scalar fields, now in the $\phi^+, \Sigma_1^+(Q_D=1), \Sigma_2^+(Q_D=-1)$ weak basis, can be written as 
\begin{equation}
{\cal M}^2_{charged}=  \begin{pmatrix}  \sqrt 2 \tilde \lambda v_tv_s & \tilde a-\tilde b &\tilde a+\tilde b\\ \tilde a-\tilde b & m_1^2 &0\\ \tilde a+\tilde b &0 & m_2^2   \end{pmatrix},
\end{equation}
where here we've now defined the abbreviations $\tilde a=\tilde \lambda v_sv_t/2$, $\tilde b=\lambda_5 v_tv_d/(2\sqrt 2)$ (with $\lambda_5 \geq 0$ assumed) as well as the two combinations 
\begin{equation} 
m_{1,2}^2 =\left( \frac{\tilde \lambda}{2\sqrt 2} \frac{v_s}{v_t} \pm \frac{\lambda_5}{4}\right)v_d^2\,,
\end{equation}
and thus we obviously must also require that $\tilde \lambda >\lambda_5 t/\sqrt 2$ in order to avoid there being a tachyonic state in the charged scalar spectrum. To leading order in $x_{t,s}$, 
$m_{1,2}$ are simply the physical masses of the fields $\Sigma_{1,2}$ which are 
split by the value of $\lambda_5 > 0$. Note the absence of any direct mixing between the $\Sigma_1^+$ and $\Sigma_2^+$ states at tree-level as they have opposite values of $Q_D=\pm 1$, 
thus differing by two units. These two new charged states will both mix with the SM $\phi^+\sim G_W^+$ by a small amount $\simeq \lambda_5 t x_t/\tilde \lambda \sim 10^{-2}$ and thus will 
not very readily couple to any of the SM fermion fields.  To this same lowest order in $x_{t,s}$ we note the `sum rule'-like relationship  
\begin{equation} 
\frac{m_H^2}{m_1^2+m_2^2}= \frac{1+t^2}{2}\,,
\end{equation}
which we might expect to be roughly $\sim O(1)$ and likely close to unity.

We will return to the detailed implications of this mass spectrum for the additional new scalars below but we can make several simple and immediate semi-quantitative (gu)estimates. 
The new, essentially degenerate, neutral fields, $H,A$, are likely to be the heaviest ones much of the time since $t+1/t \geq 2$ and so they would likely sit at the top of the mass spectrum, with 
both of the new charged states, $\Sigma_{1,2}$, lying somewhat below them (but still above the SM Higgs mass). However, we may expect that $m_2$ may frequently be as large or perhaps 
even larger than $m_H$. Clearly, for $\tilde \lambda$ fixed, assuming no other requirements, we can freely dial the quantity $t+1/t$ to a large enough value to make 
these neutral states rather heavy if we wished although here we expect $t\sim1 $. This parameter freedom is, however, absent for the case of the two charged states if no other constraints are applied. 

In order to qualitatively avoid any possible LHC constraints on the production and decay of charged Higgs pairs (which as we'll see below will for us require their masses to be 
above $\sim 230$ GeV 
or so), however, we essentially need to make the ratio $k/t$ somewhat large. For example, taking $k/t=1$ implies that that  $m_1^2+m_2^2=2v_d^2\simeq (348 ~{\rm GeV})^2$ which is not 
very large; this would seem to imply that greater values of $k/t$ will be somewhat more favored. Simultaneously, we cannot take $k$ itself too large as we need to ensure that 
 all of the quartic couplings in the potential above are perturbative, \eg, $\tilde \lambda,\lambda_5 < 4\pi \simeq 12.6$. We also cannot make $1/t=v_s/v_t$ very large as this would lead 
 to a corresponding increase in the mass 
 of the DP as described above and, if anything, we'd prefer to have $t \gsim 1$. Furthermore, we'll also simultaneously need the ratio $m_2/m_1$ sufficiently large so that, as we will see below, a 
 phenomenologically interesting value of the parameter, $\epsilon$, can be generated by KM. 

The scalar mass spectrum also roughly fixes the decay paths of these new states through both charged current and the trilinear Higgs couplings. In the charged current mode, if the 
$H,A$ are the heaviest states they will decay via (more than likely) on-shell $W$ emission to the $\Sigma_i^\pm$, which then subsequently will decay, again via on-shell $W$ emission, 
to $h_D,G_V(=V_L)$, which in turn (as we will 
assume) decay invisibly to DM thus producing missing transverse energy/momentum signatures at colliders. An alternative path, omitting the intermediate step and which depends upon the 
size of some of the scalar trilinear couplings to be discussed below, allows for the direct decay processes $H(A)\to h_{SM}h_D(G_V)$.  It is clear that final states involving $W$'s and/or 
$h_{SM}$'s plus MET will likely be the common elements in searches for the production signature for these new states.

\section{Gauge Sector Kinetic and Mass Mixing }

In the most commonly discussed PM models, KM occurs at the 1-loop level in an abelian manner between the SM $U(1)_Y$ hypercharge gauge boson and the $U(1)_D$ 
DP via a set of states having both $Y,Q_D\neq 0$. Such KM mixing is finite and, in principle, calculable in a class of models wherein the condition$\sum_i Y_iQ_{D_i}=0$ is satisfied, 
as was the case in our earlier works\cite{Rizzo:2018vlb,Rueter:2019wdf,Rueter:2020qhf,Wojcik:2020wgm,Rizzo:2021lob,Rizzo:2022qan,Wojcik:2022rtk}. In 
the present setup, the presence of the dark scalar triplet with a non-zero vev will also allow for {\it non-abelian} KM to occur between the DP and the neutral $W_3$, $SU(2)_L$ field as 
the triplet itself carries both weak isospin as well as $Q_D\neq 0$ and, in fact, alone satisfies the corresponding analogous condition $\sum_i T_{3i}Q_{D_i}=0$ with $T_3$ being the third, 
diagonal $SU(2)_L$ generator. This 
non-abelian KM can occur only after spontaneous symmetry breaking is realized as beforehand the fields in $\Sigma$ triplet will all be degenerate with 
a common mass, $m_\Sigma$, leading to the absence of KM. Of course, in a more UV-complete version of 
the present scenario, additional PM states carrying both a dark charge as well as SM hypercharge and consisting of, \eg, heavy vector-like fermions (but which are not directly part of the 
current discussion) may still be present and contribute to abelian KM. In the discussion of the current setup that follows we will generally ignore this possibility but we will 
remain mindful of its potential existence in developing our overall framework so that in the discussion that immediately follows both possibilities for KM will be considered on an equal footing. 

In general, if both types of KM between the SM fields and $U(1)_D$ are simultaneously present we can write the relevant parts of the KM Lagrangian as 
\begin{equation}
\mathcal{L}_{\textrm{KM}} = -\frac{1}{4}\hat {W}^3_{\mu \nu} \hat{W}_3^{\mu \nu} -\frac{1}{4}\hat {B}_{\mu \nu} \hat{B}^{\mu \nu} -\frac{1}{4}\hat {V}_{\mu \nu} \hat{V}^{\mu \nu} +\frac{\alpha}{2}\hat {B}_{\mu \nu} \hat{V}^{\mu \nu} +\frac{\beta}{2}\hat {W}^3_{\mu \nu} \hat{V}^{\mu \nu}\,,
\end{equation}
where the, assumed small, parameters $\alpha,\beta$, control the strength of the abelian and non-abelian KM, respectively. In the usual manner, to linear order in the $\alpha,\beta$ 
parameters and dropping the Lorentz indices for brevity, 
this generalized KM is removed by making the field redefinitions $\hat B\to B+\alpha V$, $\hat W_3\to W_3+\beta V$, and $\hat V \to V_0$. Then, employing the familiar change of basis 
$W_3=c_wZ_0+s_wA$, $B=c_wA-s_wZ_0$, (with $A$ here denoting the photon) together with making the usual identifications $e=gs_w$,  $g_Y=gt_w$,  where $s_w=\sin \theta_w$, \etc, 
we find that at this stage that the KM is indeed 
removed and that, before any mass mixing occurs, the three neutral gauge fields will now couple to the suggestive combinations (employing the definition $Q_{em}=T_3+Y/2$ as usual) 
\begin{equation}
{\cal L}_{int}=eQ_{em}A+\frac{g}{c_w}(T_3-s_w^2Q_{em})Z_0 +g(\beta ~T_3+\alpha ~t_w\frac{Y}{2})V_0\,.
\end{equation}
After $SU(2)_L$ breaking, since the $Z_0$ does not couple to the Higgs triplet vev, $v_t$, this interaction with the SM Higgs doublet alone yields the massless photon in addition to the 
$Z_0-V_0$ mass-squared matrix:
\begin{equation}
{\cal M}_{ZV}^2 = m_{Z_0}^2\begin{pmatrix} 1 & c_w\beta-s_w\alpha \\ c_w\beta-s_w\alpha & (c_w\beta-s_w\alpha)^2+\gamma^2\end{pmatrix},
\end{equation}
where $m_{Z_0}=gv_d/(2c_w)$ is the `SM' $Z$ mass and $\gamma^2=m_{V_0}^2/m_{Z_0}^2<<1$ with $m_{V_0}$ as given above{\footnote {We also note the obvious result 
that one finds $m_W^2=c_w^2m_{Z_0}^2(1+4x_t^2)$.}}.

This matrix can be diagonalized via the small rotation $Z_0\simeq Z-\chi V$, $V_0\simeq V+\chi Z$ with the angle $\chi \simeq s_w\alpha-c_w\beta$ assuming that $|\chi|<<1$. 
To lowest order in $\alpha, \beta$ and $\gamma^2$ one find that the $V,Z$ mass eigenstates maintain
the same masses as was had by $V_0$ and $Z_0$, the $Z$ also maintains its familiar SM coupling structure a la the $Z_0$ above and, taking as is conventional\cite{vectorportal} 
$\alpha \to \tilde \alpha/c_w$, $\beta \to \tilde \beta/s_w$, one finds that the physical DP, $V$, will couple to the SM fields as 
\begin{equation}
{\cal L}_{V_SM} \simeq e \left(\tilde \alpha+ \tilde \beta\right)Q_{em}~V = e\epsilon_{eff}Q~V\,,
\end{equation}
as might be expected. Note that $\epsilon_{eff}$ can in principle receive comparable contributions from both abelian and non-abelian KM sources or only one of these may be dominant. 
The present model as currently described 
has no PM fields with $Y\neq 0$ so that all of the KM is non-abelian in origin via the dark isotriplet $\Sigma$ but the effect of this mixing yields that same type of DP coupling to SM matter 
fields, in the parameter space region of interest to us here, as does the more conventional abelian KM. 

The new scalar fields we have introduced are now seen to generate a contribution to KM via the familiar set of 1-loop graphs. Since the SM $Z$ does not couple to $S$ and, in fact, to none 
of the neutral scalar fields which all have $T_3=0$, only the newly introduced charged states with both $Q_D=1$ and $T_3=\pm 1$ can contribute here. From these 1-loop graphs 
only involving these charged $\Sigma_{1,2}^\pm$ fields, we find the finite and somewhat familiar-looking result
\begin{equation} 
\epsilon_\Sigma= \left(\frac{g_D}{e}\right)\cdot \frac{\alpha_{QED}}{12 \pi}\textrm{ln} \left(\frac{m_2^2}{m_1^2} \right).
\end{equation}
Recall, as noted above, that before SSB, when all of the various vevs get turned on, the $\Sigma$ triplet is degenerate so that this source of KM would be turned off. 
Given the expressions for the masses $m^2_{1,2}$ above one can immediately calculate this quantity in any specific model as it depends only upon the parameters $t,\tilde \lambda$ and 
$\lambda_5$  in the form of the simple ratio, $z=\lambda_5 t/(\sqrt 2\tilde \lambda)$, as always, to lowest order in the small parameters, $x_{s,t}$. Numerically, we see that 
if, \eg,  $m_2=1.5(2)m_1$, then one finds that 
$\epsilon_\Sigma= 1.7(2.9)~\left[\frac{g_D}{e}\right] \cdot 10^{-4}$, values that are phenomenologically interesting\cite{Battaglieri:2017aum}; to obtain these mass ratios requires that 
$z\simeq 0.38(0.60)$, respectively. We also see that, \eg, if $z$ is less 
than $\simeq 0.25$, then $\epsilon_\Sigma$ will lie (if $g_D=e$) below $10^{-4}$ and that other KM sources will then likely be needed for a DP in the mass range of interest here so that the 
range of this parameter is then somewhat restricted. Of course, as previously noted, in a more UV-complete framework beyond the current setup, 
other new fields, \eg, in the form of $Q_D \neq 0$ heavy 
vector-like fermions\cite{Rizzo:2018vlb,Rueter:2019wdf,Kim:2019oyh,Rueter:2020qhf,Wojcik:2020wgm,Rizzo:2021lob,Rizzo:2022qan,Wojcik:2022rtk}, are also potentially present, carrying 
non-zero values the SM hypercharge so that additional contributions to KM from this more `conventional' abelian source are obtainable. Here, for simplicity, we will ignore this possibility 
assuming the $\epsilon_\Sigma$ is the only source of KM. 

Interestingly,  now that the masses of the various gauge fields are fully determined, we can symbolically relate the apparent shift in the $W$ mass away 
from the expectations of the SM, as was measured by CDFII, directly with our model parameters as employed above; specifically, we find that 
\begin{equation}
\frac{\Delta m_W^2}{m_W^2}=\frac{4(1-s^2_w)}{1-2s^2_w}~\frac{t^2}{g_D^2(1+t^2)v_d^2}~m_V^2,
\end{equation}
which is a rather amusing result.

At this point it is instructive to examine what the (tree-level) effects of these new scalars and their vevs may be on, \eg,  the gauge boson partial widths of the SM-like Higgs state at $\sim 125$ 
GeV as observed at the LHC{\footnote{Note that the $Z$ boson and SM fermion couplings are modified in an identical manner in the present setup.}}. This may best be analyzed by way of the 
familiar $\kappa$ rescaling parameters used to describe the deviations of Higgs couplings/partial widths from SM expectations as are recently summarized in 
Refs.\cite{ATLAS:2022vkf,CMS:2022dwd}. Using the physical (\ie, measured) $W,Z$ masses as input (and to avoid any possible renormalization scheme ambiguities), it is useful to examine 
the (double) ratio of Higgs partial widths $R=R_{WZ}/R_{WZ}^{SM}=[\Gamma(WW^*)/\Gamma(ZZ^*)]/\rm{SM}=[\kappa_W/\kappa_Z]^2$ in comparison to SM expectations, again, here all 
at the {\it tree-level}. Let us denote the $h_{SM}(=h_{125})$ content of the CP-even weak eigenstate fields $\phi,s,\sigma$ by $f_{\phi,s,\sigma}$ as can be obtained via the diagonalization of 
the corresponding CP-even Higgs mass matrix given above; note that whereas one finds that $f_\phi \simeq 1$, we instead find that $f_s\sim O(x_{s,t})$ and more than likely even 
somewhat smaller in the case of $f_\sigma$. Some algebra then tells us that $R$ is just given by the ratio 
\begin{equation}
R= \Big[\frac{1+4x_tf_\sigma/f_\phi}{1+\delta \rho}\Big]^2\simeq 1+8x_t(f_\sigma-x_t),
\end{equation}
so that we may expect that the numerical value of $R$ will deviate from unity by terms of order $O(x_{t,s}^2)< 10^{-3}$, a shift which is likely far too small as to be accessible in the 
foreseeable future, even at proposed lepton and hadron colliders\cite{deBlas:2022ofj}. We caution the Reader, however, that due to the very small size of these apparent shifts obtained 
at the tree-level, loop-order radiative corrections in this model must be included to ascertain the importance of their potential influence on the expected value of this ratio before any 
comparison with experiments can be made.

\section{New Scalar Production and Decay Phenomenology}

\begin{figure}
\centerline{\includegraphics[width=5.0in,angle=0]{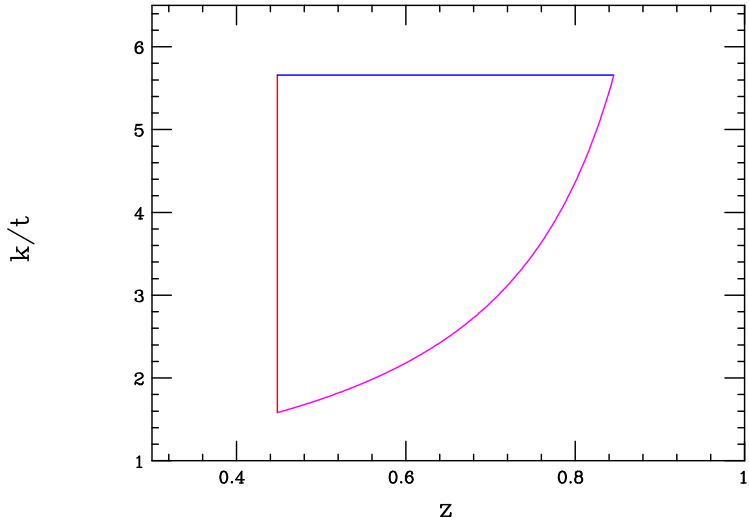}}
\vspace*{-1.10cm}
\caption{The constraints on the model parameter space in the $k/t-z$ plane as given and described in the discussion in the text. The left (red) boundary is set by the lower bound on 
$\epsilon_\Sigma$ while the top (blue) boundary by the upper bound on $\tilde \lambda/t$. The lower (magenta) boundary is set by the minimum value of $m_1$. The allowed region 
then lies within all of the curves.}
\label{fig0}
\end{figure}

\begin{figure}[htbp]
\centerline{\includegraphics[width=4.1in,angle=0]{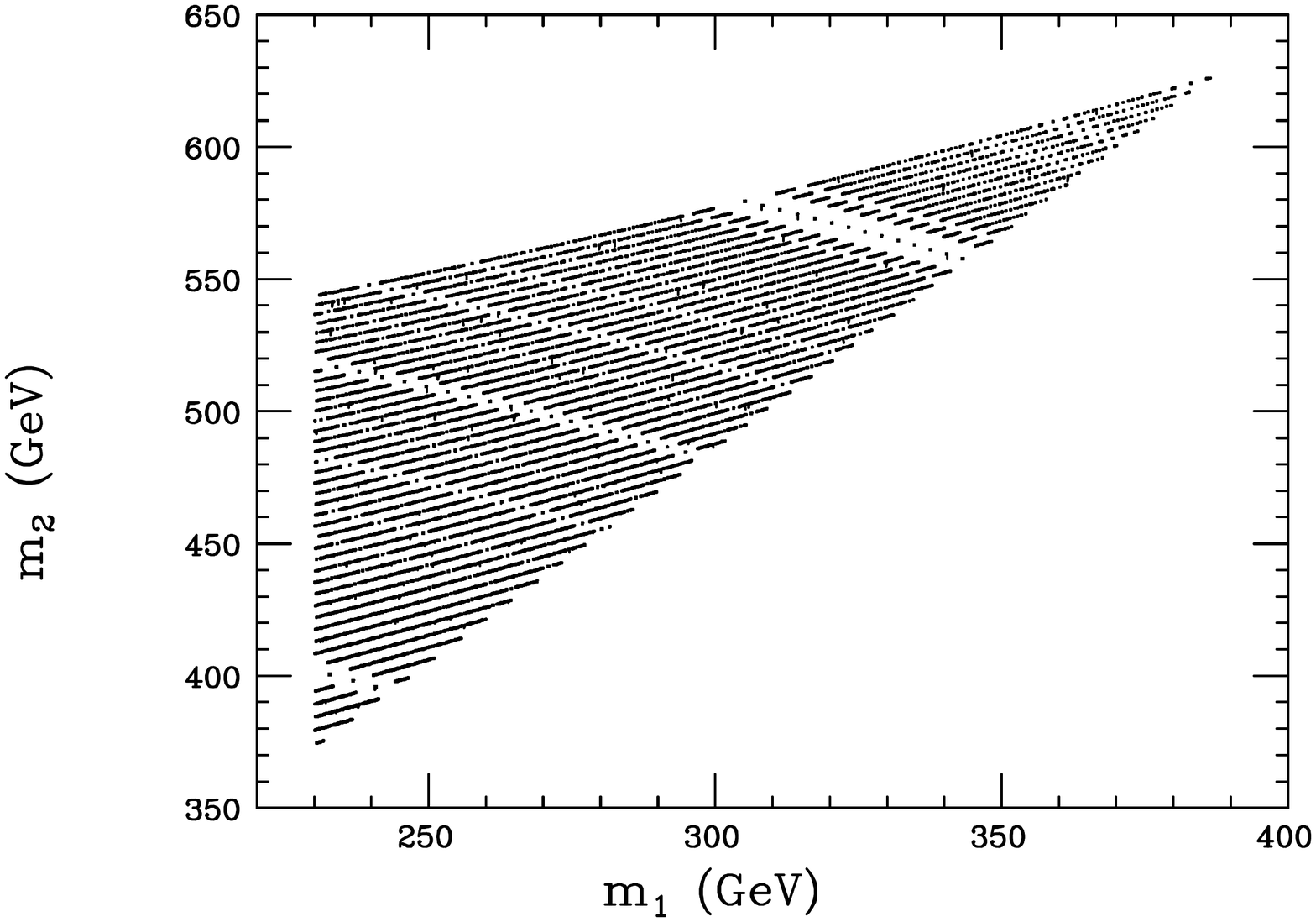}
\hspace*{-1.7cm}
\includegraphics[width=4.1in,angle=0]{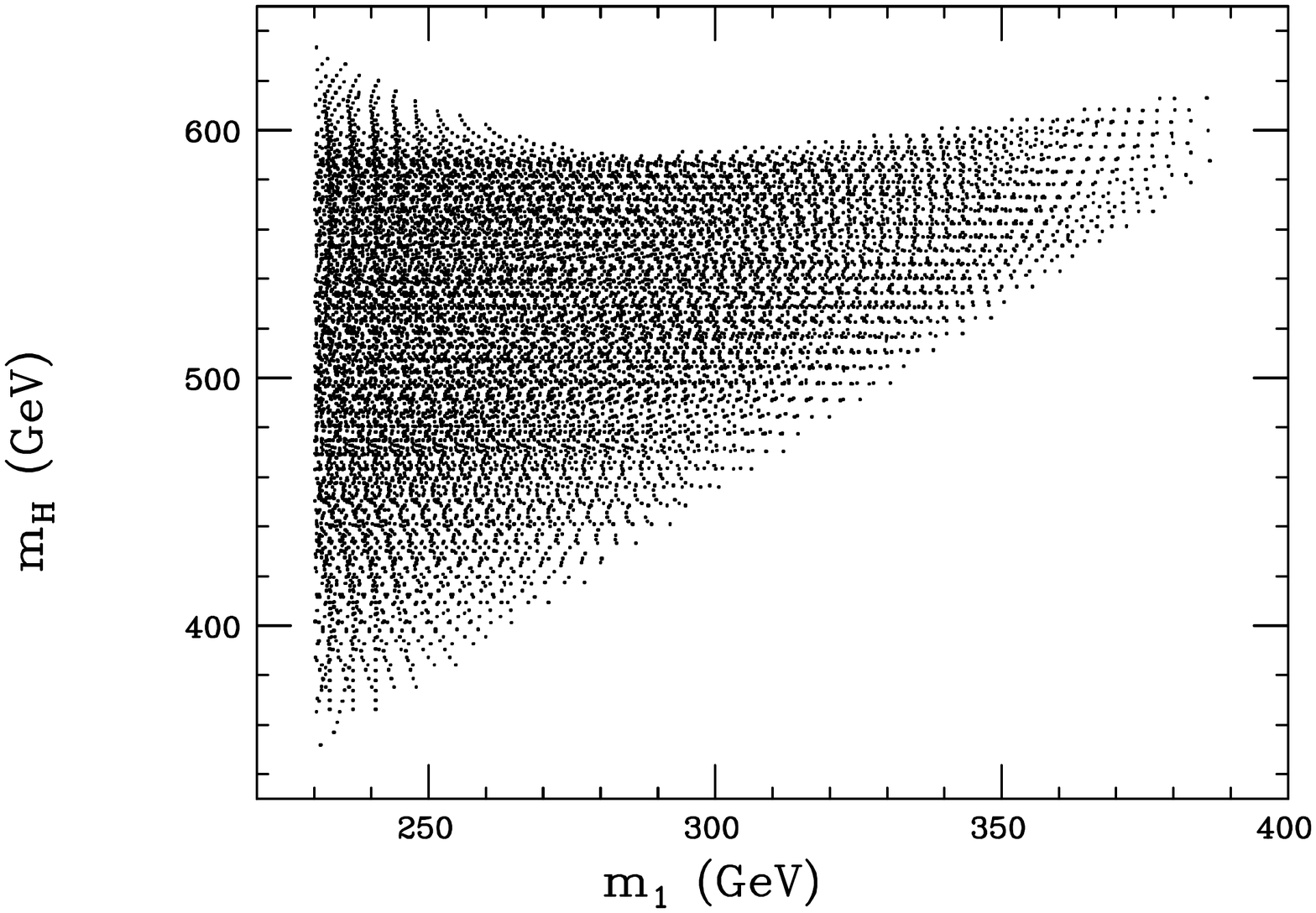}}
\vspace*{-1.4cm}
\centerline{\includegraphics[width=4.1in,angle=0]{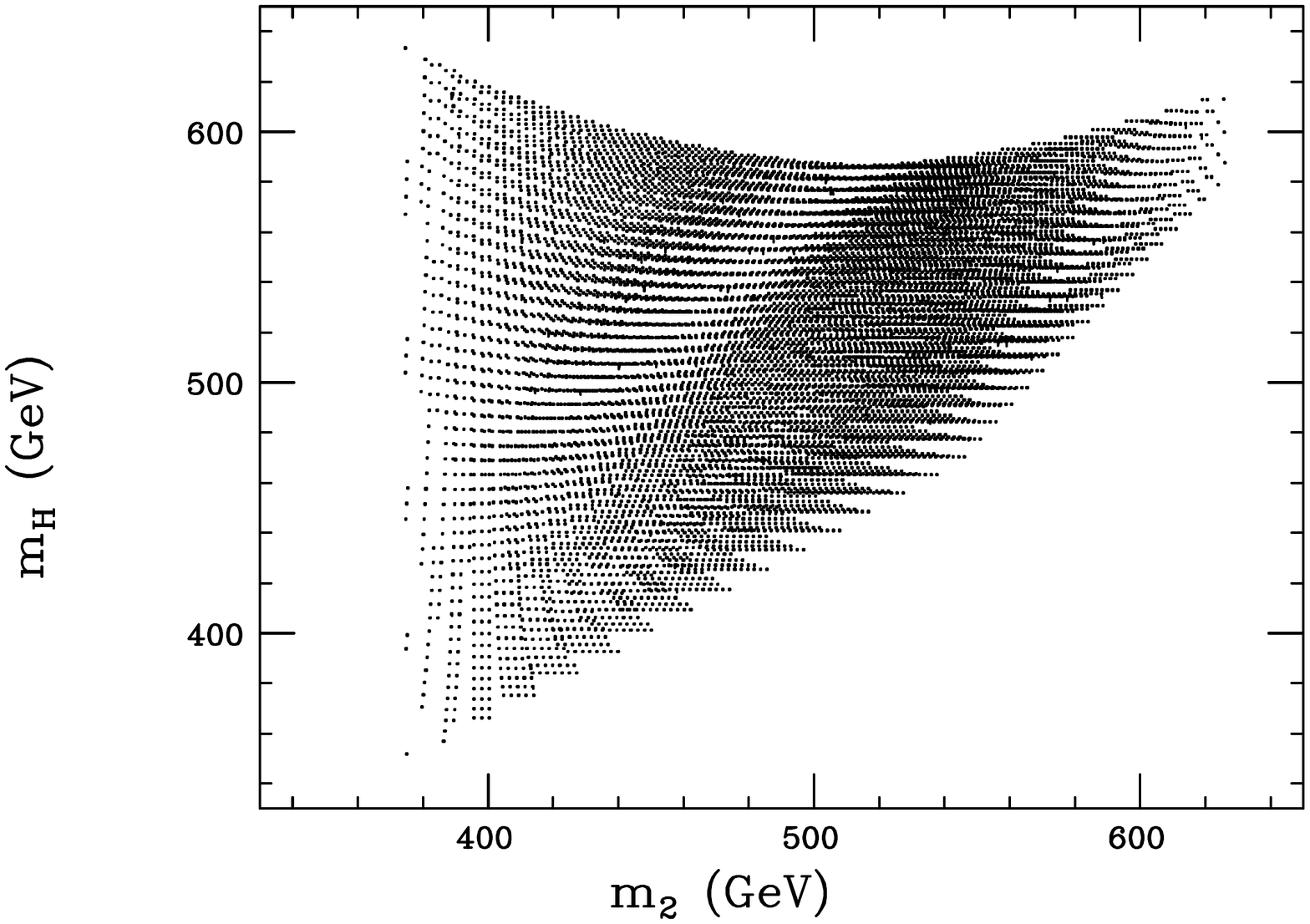}}
\vspace*{-1.20cm}
\caption{ Results for the heavy scalar PM mass spectra arising from the coarse-grained scan of parameters subject to the assumptions and 
constraints as described in the text. (Top Left) $m_2-m_1$ projection, (Top Left) $m_H-m_1$ projection, (Bottom) $m_H-m_2$ projection.}
\label{scan}
\end{figure}

Before turning to the specifics of phenomenology, it may be useful to perform a modest scan of the most relevant parts of the model parameter space, consisting of the set 
$(\tilde \lambda, t, \lambda_5)$, 
to see where the preferred regions may be after applying some simple selection cuts as discussed above.  In this flat parameter scan, we consider as an example the ranges 
$0.5\leq t \leq 2$ and $1\leq \lambda_5,\tilde \lambda \leq 8$ (\ie, safely away from the perturbativity bound), together with the two requirements that 
($i$) $\epsilon_\Sigma \geq 2(g_D/e)\cdot 10^{-4}$ under the assumption that no other additional PM states are present and also ($ii$) $m_1\geq 230$ GeV (to cleanly avoid constraints from the LHC as we'll see below). The first constraint, ($i$), immediately leads to a lower limit on the value of the parameter $z$ introduced and discussed above, \ie, 
$z=\lambda_5t/\sqrt 2 \tilde \lambda \gsim 0.45=z_{min}$.  This value of $z_{min}$ implies that $m_2/m_1 \geq 1.62$, a result which will impact our discussion below. 
In terms of the parameter $k=\tilde \lambda/(2\sqrt 2)$, also introduced above, since $m_1^2=k(1-z)v_d^2/t$, requirement $(ii$) 
leads to bound $k/t \gsim1.87$ and, using the previously obtained value of $z_{min}$, one thus finds $4.47 t\lsim \tilde \lambda \leq 8$, hence that $t\lsim 1.79$. Now since both $0.5\lsim t$ and 
$\tilde \lambda \lsim 8$, one obtains $k/t \lsim 5.66$ and, hence, $z$ is also bounded from above, $z \lsim 0.85=z_{max}$, also implying that $\lambda_5\lsim 5.35$. The results of 
these straightforward considerations for the $k/t-z$ parameter plane are shown in Fig.~\ref{fig0}. In addition to these parameter constraints, and perhaps even more importantly, the expected 
scalar PM mass ranges are then also found to be restricted: $230 \lsim m_1\lsim 385$ GeV,  $360\lsim m_2\lsim 630$ GeV and $360 \lsim m_{H,A}\lsim 640$ GeV, with some obvious 
correlations between these separate ranges of values to be expected.  The results of a rather coarse-grained grid scan over the model parameters satisfying the two constraints above are shown 
in Fig.~\ref{scan} where all the correlations are plainly visible. 
These numerical results will have some important influence on the discussion of the production processes and decay signatures for the new heavy scalars which now follows. 


To begin our discussion of the phenomenology of the new heavy PM scalars in our setup, we first consider how these states may be produced at the LHC as well as their subsequent decays in 
order to access their detectability. Clearly the gauge interactions of these states as well as their trilinear self-interactions, particularly with $h_{SM}$ as those are the ones proportional to 
the large vev, $v_d$,  will play the dominant role here. Such trilinear interactions are described by the following part of the full model Lagrangian:
\begin{equation}
\begin{aligned}
\mathcal{L} \supset ~\frac{h_{SM} v_d}{2} &\bigg\{ (h_D^2 + G_V^2) \left[\frac{t^2}{2(1+t^2)}(\lambda_{4\Phi}+\lambda_5)+\frac{\kappa}{1+t^2}-\frac{\sqrt 2 \tilde \lambda ~t}{1+t^2}\right]  \\
&+(A G_V+H h_d) \left[ \frac{t}{1+t^2}(\lambda_{4\Phi}+\lambda_5-2\kappa)-\sqrt 2 \tilde \lambda ~\frac{1-t^2}{1+t^2}\right] \\
&+(H^2 + A^2) \left[ \frac{1}{2(1+t^2)}(\lambda_{4\Phi}+\lambda_5)+\frac{\kappa t^2}{1+t^2}+\frac{\sqrt 2 \tilde \lambda ~t}{1+t^2}\right]\\ 
&+ 2\lambda_{4\Phi} \Sigma_1^+\Sigma_1^- +2(\lambda_{4\Phi}+\lambda_5)\Sigma_2^+\Sigma_2^- \bigg\}.
\end{aligned} 
\end{equation}
This piece of the scalar Lagrangian controls a multitude of interesting interactions that we will examine below; other possible trilinear couplings are found to be relatively suppressed 
by at least one power of the small ratios $x_{t,s}$. For example, the cross section for resonant SM di-Higgs production via the heavy $H$ is highly suppressed in this setup by at 
least four powers of $x_{t,s}$.

As was previously mentioned, one general difference between the current setup and the two dark doublet PM model examined earlier\cite{Rueter:2020qhf} 
and which will occur as a continuing theme in the discussion below is the fact that 
all of the neutral scalar fields here have $T_3=0$ and so do not couple to the SM $Z$. This not only eliminates potential production mechanisms for these new scalars but also many possible decay  
paths which results in far fewer new collider signatures here by comparison to this earlier examined scenario. This results in fewer LHC analyses constraining the present setup by comparison 
to the two doublet case which relied heavily on the clean final states initiated by, \eg,  the leptonic decays of the $Z$. 

We first turn to the production of these new scalar particles at the LHC via off-shell SM gauge boson exchange in the $s$-channel from initial $\bar q q$ states. $\gamma^*,Z^*$ exchange 
can lead to $\Sigma_i^+ \Sigma_i^-$ pair production with cross sections totally fixed by their masses and electroweak couplings. If the hierarchy between the masses $m_{1,2}$ is 
sufficiently large, 
this rate will be almost totally dominated by $\Sigma_1$ pairs. On the other hand, $W^*$ exchange can lead, to lowest order in the small mixing angles, to the eight possible final states 
$\Sigma_{1,2}\times (H,h_D,A,G_V)$ with the $\Sigma_1$ contribution being kinematically dominant and with production rates being fixed by the particle masses and the value of $t$ which 
determines the mixing angle factor appearing in these couplings. Due to the approximate mass degeneracies above, the production rates for the two choices $H/A$ and $h_D/G_V$ of 
finals states are found to be the same with the former pair being relatively kinematically suppressed due to the significantly larger values of the $H,A$ masses as was discussed earlier. 
Note, however, that unlike in the case of dark scalar doublet PM model\cite{Rueter:2020qhf}, there is in the present setup no analogous $Z^*$ exchange process leading to the 
four purely neutral $(H,h_D)\times (A,G_V)$ final states since, to lowest order in the $x_{s,t}$'s, all of the new neutral scalar states have $T_3=0$ and so do not couple to the SM $Z$. 

\begin{figure}
\centerline{\includegraphics[width=5.0in,angle=0]{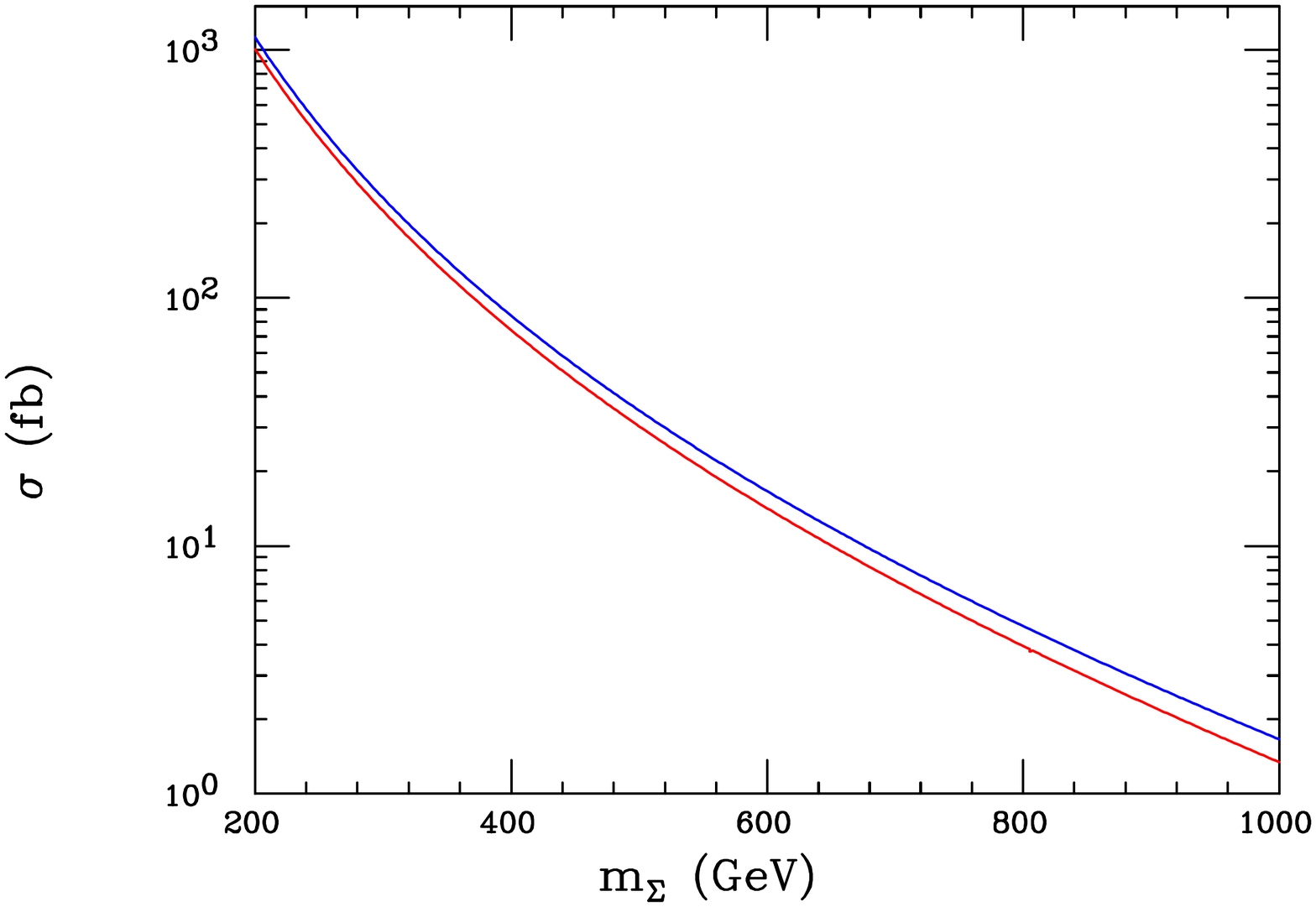}}
\vspace*{-2.4cm}
\centerline{\includegraphics[width=5.0in,angle=0]{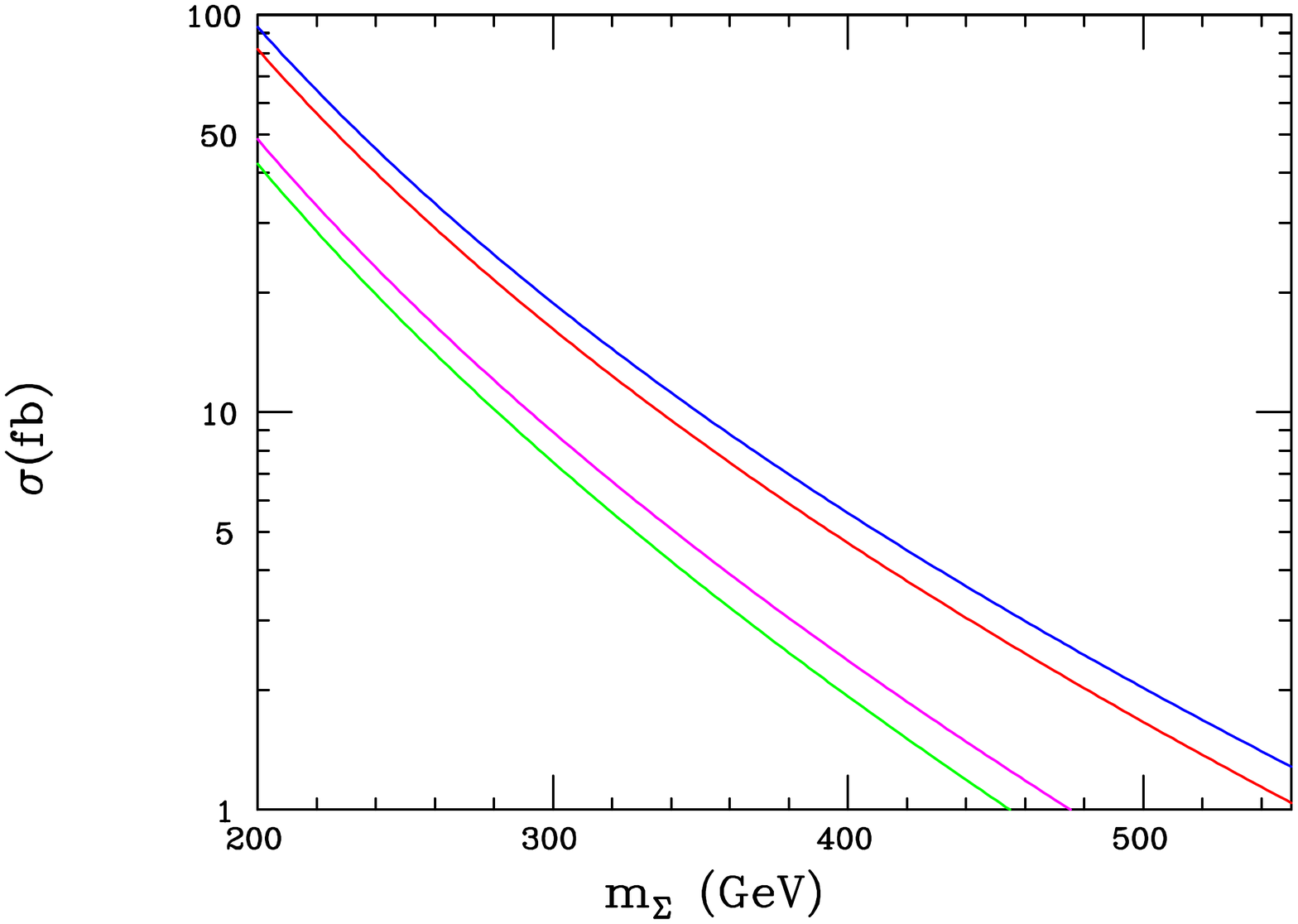}}
\vspace*{-1.30cm}
\caption{(Top) The cross section for $\Sigma_i^\pm (h_D+G_V)$ production via $W^*$-exchange as a function of $m_\Sigma$ at the LHC assuming $\sqrt s=13$ (red) or 14 (blue) TeV 
and assuming that $m_{h_D, V}=0$ and $s^2_\phi=1$. (Bottom) Cross sections for $\Sigma^+\Sigma^-$ pair production via $\gamma^*,Z^*$, $s$-channel exchange at the $\sqrt s=13 (14)$ TeV LHC 
in red (blue) as a function of $m_\Sigma$. The corresponding cross sections, with $c^2_\phi=1$, for $\Sigma_1^\pm (H+A)$ production via $W^*$ $s-$channel exchange assuming 
$\sqrt s=13 (14)$ TeV at the LHC in green (magenta) are also shown. Here the representative mass relation $m_{H,A}=1.8m_{\Sigma_1}$ has been employed for purposes of demonstration.}
\label{fig1}
\end{figure}

To get a feel for the rates for these processes, the top panel in Fig.~\ref{fig1} shows the virtual $W^\pm$-exchange production cross section for the sum of the two final states 
$\Sigma_i h_D+\Sigma_i G_V$ as a function of the $\Sigma_i$ mass taking $m_{h_D,V}=0$ when $s^2_\phi=1$ is assumed; for example, when $t=1$ the cross section shown 
would need to 
be reduced by a factor of 2. This is the largest  - and hence most important - cross section associated with the production of the new scalar states since, neglecting the $O(1)$ mixing angle 
factors themselves, ($i$) $W^*$-induced processes are larger due to the presence of somewhat larger weak coupling factors and ($ii$) only one heavy scalar is produced in the final state 
which reduces the overall phase space penalty.  The lower panel in Fig.~\ref{fig1} shows that the cross sections for charged $\Sigma^+_i \Sigma^-_i$ pair production and 
also for $\Sigma_1H+\Sigma_1A$ associated production (here under the assumptions that $c^2_\phi=1$ and with $m_{H,A}=1.8m_{\Sigma_1}$) are very roughly comparable (although 
the rate for the associated production mechanism is a bit smaller by a factor of $\sim 2$) and are suppressed in comparison to that for 
 $\Sigma_1h_D+\Sigma_1G_V$ by roughly factors of order $\sim 20-30$ or more. 

\begin{figure}
\centerline{\includegraphics[width=5.0in,angle=0]{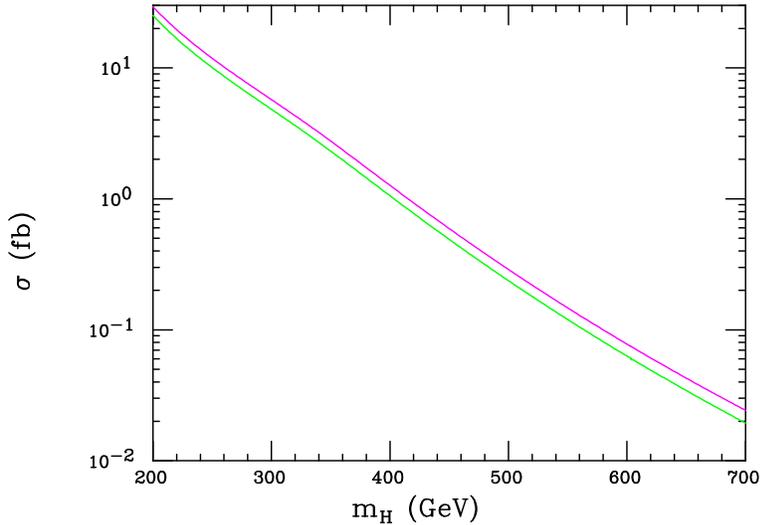}}
\vspace*{-1.10cm}
\caption{The $gg\to h_{SM}^* \to Hh_D+AG_V $ cross section as a function of $m_{H,A}$ at the $\sqrt s=13 (14)$ TeV LHC in green (magenta) assuming that the trilinear $Hh_Dh_{SM}$ 
coupling, as defined in the text, $|\lambda''|=1$. Here it is again assumed that $m_{h_D, V}=0$.}
\label{fig2}
\end{figure}

A third collider production mechanism is via $s$-channel, off-shell SM Higgs exchange, \ie, $gg\to h_{SM}^* \to Hh_D,AG_V$, which takes place through the usual triangle graph and 
with a cross section quadratically dependent upon the {\it a priori} unknown size of the trilinear, \eg, $Hh_Dh_{SM}$ coupling given in the $\mathcal{L}$ expression above. We will describe this 
interaction by the effective vertex $v_d\lambda''/2$ with $\lambda''$ being the expression inside the relevant square bracket and it is this same combination which controls the decay rates for  
$H(A)\to h_{SM}h_D(G_V)$ as briefly discussed above. However, we note that even when we assume that $|\lambda''| \sim 1$, this 
cross section is found to be rather small especially if we expect that the neutral scalars themselves masses to be somewhat large, \eg, $m_{H,A} \gsim 360$ GeV, as can be 
seen in Fig.~\ref{fig2}. Note that there are, \eg, 
no corresponding $H$ exchange graphs as, to lowest order in $x_{t,s}$, $H$ does not couple to the SM fermions and so no corresponding $ggH$-type coupling can be generated.

\begin{figure}
\centerline{\includegraphics[width=5.0in,angle=0]{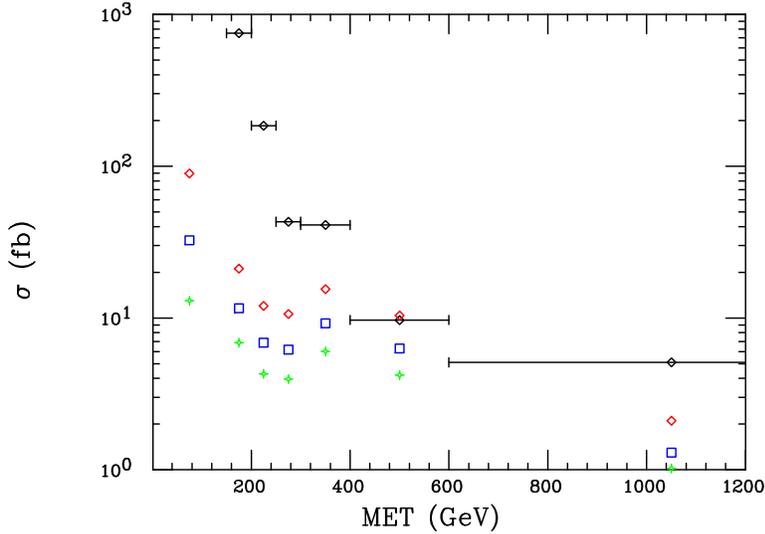}}
\vspace*{-1.10cm}
\caption{Comparison of the 95$\%$ CL upper bounds obtained on the production of the $W^\pm+$MET signal in the all-hadronic channel (solid, also showing the various bin sizes) obtained 
by ATLAS\cite{ATLAS:2018nda} for different MET bins in comparison with the expectations of the current setup with $s_\phi^2=1$ and $m_1=200$ (red diamonds), 250 (blue squares) or 300 
(green crosses) GeV. }
\label{fig3}
\end{figure}

From this discussion, it is clear that $\Sigma_1$ pair production and $\Sigma_1+h_D/V(=G_V)$ associated production will likely supply the most important signals and 
constraints on the present model as 
the $\Sigma_1 +H/A$ process is significantly suppressed by the heavy particle phase space and also generally produces a more complex final state to reconstruct. The former process 
leads to the final state of $W^+W^-+$MET while the later one simply to $W^\pm+$MET. The constraints on $\Sigma_1$ pair production can be approximately obtained by 
recasting the searches for SUSY wino-like chargino pairs (which it closely resembles), assuming a 100$\%$ branching fraction for the $\tilde \chi^\pm \to W^\pm+$MET decay mode, 
correcting for the differences in the production process. Here we make specific use of the wino-like chargino pair production, dilepton+MET analysis at $\sqrt s=13$ TeV with an 
integrated luminosity of $L=139$ fb$^{-1}$ as was performed by ATLAS\cite{ATLAS:2019lff} in the limit of a massless LSP{\footnote {The corresponding $W$ all-hadronic 
decay analysis from CMS\cite{CMS:2022sfi} has little impact here due to the smaller 
cross sections found for scalars in comparison to those for wino-like charginos. As noted by CMS, Higgsino-like charginos are also unconstrained by their search.}}. From this analyses one 
finds that we must roughly have $m_1 \gsim 230$ GeV. To obtain this result we have assumed that pairs of $\Sigma_1$'s 
will completely dominate in the production of this signal and that $B(\Sigma_1 \to W+$MET$)=1$. Of course, in reality, $\Sigma_2$ pairs are also produced but with a suppressed rate 
which depends upon the the mass splitting 
between these two states. However, we know that $m_2/m_1\geq 1.62$ from the discussion above while the parameter scan for low $m_1$ tell us this ratio can be as large as 2.39.  Given 
the mass dependence of the cross section shown in Fig.~\ref{fig3} it is thus a reasonable approximation to neglect the pair production of $\Sigma_2$'s. It should be noted that, in comparison to the 
fermionic (\ie, wino-like chargino) case, the present scalar scenario suffers not only from a suppressed cross section but also 
a slower turn-on at threshold behavior, both of which somewhat reduce the efficacy of the experimental cuts.

Something similar happens in the case of $\Sigma_1+h_D/G_V$ associated production where $\Sigma_1$ then decays to $W+$MET; here we make use of the 13 TeV, 36.1 fb$^{-1}$ 
ATLAS analysis\cite{ATLAS:2018nda} with the $W$ decaying hadronically. However, in this case the cuts are not so well designed for our particular kinematics but we can still 
perform a recasting of their results. Effectively, here one finds that 
when $\Sigma_1$ is light, thus yielding a large cross section, it is difficult for events to pass the MET requirements; when $\Sigma_1$ is more massive, thus yielding a larger MET in the final 
state, the cross section is suppressed. This was found to be a common feature of the parallel analyses performed earlier in the case of scalar PM doublets\cite{Rueter:2020qhf}. 
Fig.~\ref{fig3} compares the effective cross section upper bound for this final state as obtained by ATLAS for several different MET bins with the predictions of the present model 
assuming that $s_\phi^2=1$ and employing 
different values of $m_1$. In this analysis, we have as before assumed that $\Sigma_2$ is too massive to make any substantial contribution to the overall event rate which, given 
the fall off of the cross sections with mass as seen in Fig.~\ref{fig1} seems reasonable.  Taken literally, 
from this Figure, we see that the predicted cross section in the 500 GeV MET bin exceeds the ATLAS bound when $m_1=200$ GeV is assumed, excluding this possibility. However, increasing 
the mass, even to 250 GeV, is seen to allow us to survive this particular constraint although we can easily imagine that by increasing this search's integrated luminosity to $\simeq 139$ fb$^{-1}$ 
might possibly exclude this mass value (still assuming $s_\phi^2=1$) but certainly not a much larger one. Thus the previous ATLAS bound of $\simeq 230$ GeV obtained from the 
$W^+W^-+MET$ analysis 
discussed above is essentially reproduced in this analysis and so still appears as the most realistic one currently without a more detailed study. Employing similar, but possibly smarter 
analyses in various combinations and moving to $\sim 14$ TeV at the HL-LHC 
may allow us to probe masses up to roughly $m_1 \sim 400-450$ GeV, again under the assumption that $s_\phi^2=1$.  

Allowing for the variations in $t$ above and by combining both of these MET analyses it is more than likely that the HL-LHC will be able to cover most if not completely all of the 
allowed parameter space for this scenario. 

Some of the lighter of the new fields in this model might also be produced in the decays of SM particles, \eg, $Z\to Vh_D$ (\ie, $G_Vh_D$), which contributes to the $Z\to {\rm invisible}$ 
signal,  and, analogously, $h_{SM} \to 2h_D,2G_V$, which leads to $h_{SM}\to {\rm invisible}$,  as was mentioned above. However, unlike in the PM scalar doublet 
scenario\cite{Rueter:2020qhf}, all of the new neutral scalar fields have $T_3=0$ to leading order in $x_{s,t}$ so that the former process is highly suppressed to an uninteresting 
level. The later process, however, remains a strong constraint on 
the model parameter space via the effective $h_{SM}h_D^2$ interaction vertex, which we here denote as $ v_d\lambda'/2$, and which appears in the $\mathcal{L}$ trilinear scalar coupling 
expression above. $\lambda'$, like $\lambda''$ encountered earlier, is seen to be a function of the 5 parameters $\kappa, t, \lambda_{4\Phi,5}$ and $\tilde \lambda$ once higher order terms 
in $x_{t,x}$ are neglected. This interaction leads to the partial decay width (taking the the limit where both the dark Higgs and dark photon masses are set to zero) given by
\begin{equation}
\Gamma(h_{SM} \rightarrow 2h_D,~2G_V) = \frac{(\lambda' v_d)^2}{32 \pi m_{h_{SM}}},
\end{equation}
which, as noted earlier, must yield a branching fraction satisfying the experimental bound $B(h_{SM} \to {\rm invisible}) <0.11$\cite{ATLAS:2020kdi} and this leads to the rather 
stringent and highly 
tuned constraint on this combination of parameters, \ie,  $|\lambda'| \lsim 6.5\cdot 10^{-3}$, quite similar to what happened in the case of the two scalar PM doublets\cite{Rueter:2020qhf}. Given 
previously discussed constraints on $t$, $\lambda_5$ and $\tilde \lambda$, this highly restricts the remaining so far free parameters, $\kappa$ and $\lambda_{4\Phi}$.  It is interesting 
to re-write this effective coupling, $\lambda'$,  in 
terms of the $\Sigma_i$ masses and the remaining model parameters, noting the potential cancellation that must occur between the positive and negative contributions to realize 
this rather small value as 
\begin{equation}
\lambda'=\frac{t^2}{1+t^2} \left(\frac{\lambda_{4\Phi}}{2} +\frac{\kappa}{t^2}-\frac{m_2^2+3m_1^2}{v_d^2}\right),
\end{equation}
clearly showing that smaller values of the  $\Sigma_{1,2}$ masses are somewhat preferred by this constraint.

Another potentially interesting process, $h_{SM}\to ZVh_D \to Z+$MET can happen at tree-level in the PM doublet model but is found to be 
absent here to lowest order in $x_{t,s}$ since the neutral 
Higgs scalars do not couple to the $Z$ in the current scenario. At the loop-level, the process $h_{SM}\to ZV, V\to$ MET can occur but it is found to be even more suppressed than the 
process to which we now turn.

\begin{figure}
\centerline{\includegraphics[width=5.0in,angle=0]{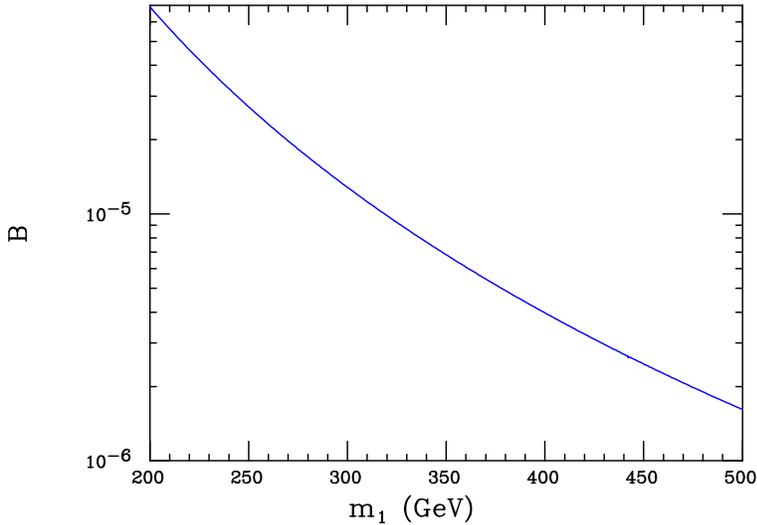}}
\vspace*{-1.10cm}
\caption{Branching fraction for the process $B(h_{SM}\to \gamma V)$ as a function of the mass of the lightest charged scalar, $m_1$ assuming $g_D=e$ and $c_1=1$ in the limit that 
the destructive contribution from the more massive $\Sigma_2$ charged scalar can be neglected.}
\label{fig4}
\end{figure}

As is well-known, and as we'll see in a bit more detail below, the influence of charged Higgs states above the top quark mass on loop-induced SM Higgs decays, such as 
$h_{SM}\to Z\gamma,2\gamma$ are relatively weak\cite{Gunion:1989we}.   
However, since these fields in the present setup also carry dark charges, the decay $h_{SM}\to \gamma V, V\to$ MET signal can be induced by these same (albeit now destructively interfering) 
$\Sigma_i$ loops as was the case in the previously examined PM doublet scenario\cite{Rueter:2020qhf}. In the SM, this same $\gamma +$MET final state can also be achieved via the usual 
$h_{SM}\to \gamma Z, Z\to \bar \nu \nu$ process which has a branching fraction of roughly $\simeq 3 \cdot 10^{-4}$ and which now forms an irreducible background to the current reaction 
under investigation.  For brevity,  we here denote the $h_{SM}\Sigma_i^+\Sigma_i^-$ couplings as $c_i v_d$ with $c_1=\lambda_{4\Phi}$ 
and $c_2=\lambda_{4\Phi}+\lambda_5$ as can be seen the Lagrangian above. Note that as $\lambda_5\to 0$, these two charged fields will become degenerate and will also have the 
same coupling to $h_{SM}$ 
thus producing complete destructive interference as would be expected; the decay $h_{SM}\to VZ$ experiences the same sort of destructive interference since the $\Sigma_i$ have opposite 
$T_3$ values as well opposite values of $Q_{em}$. Following Ref.\cite{Rueter:2020qhf}, and defining $\tau_i=4m_i^2/m_{SM}^2$, we may express the branching fraction for this process as 
\begin{equation}
B(h_{SM}\to \gamma V) \simeq 0.23~ \frac{g_D^2}{e^2}~ \Big (c_1F(\tau_1)-c_2F(\tau_2)\Big)^2\\,
\end{equation}
where we have defined the familiar loop function, $F$, here to be 
\begin{equation} 
F(\tau)= -\frac{1}{2} \Big(1-\tau[\sin^{-1} (1/\sqrt{\tau})]^2\Big)\simeq \frac{1}{6\tau}\,. 
\end{equation}
Unfortunately, Fig.~\ref{fig4} shows the branching fraction for the process $B(h_{SM}\to \gamma V)$ as a function of $m_1$ with both $g_D/e$ and $c_1$ set 
to unity and neglecting any of the effects of the 
destructive interference arising from the heavier $\Sigma_2$. Here we see that any potential signal lies at least an order of magnitude below the expected SM background even when these 
destructive interference contributions are neglected.

Finally, as mentioned above, we consider the influence of the now constructively interfering, new charged Higgs states, $\Sigma_i^\pm$, on the partial width $h_{SM}\to 2\gamma$ which 
is dominated by the $W$ and top loop contributions in the SM. Using the couplings as given in Eq.(23) above as well as the loop functions and other machinery from, \eg, 
Ref.\cite{Gunion:1989we}, we can estimate the fractional shift,  $\Delta$,  in this partial width to leading order as a function of the $c_i$ assuming some input values for the charge Higgs 
masses, $m_i$. From Fig.~\ref{scan} above, the masses $m_1=275$ GeV and $m_2=500$ GeV are quite suggestive and a short calculation assuming these values then yields the result that 
\begin{equation} 
\Delta\simeq \Big( 2.06~\frac{c_1}{g^2}+0.534~\frac{c_2}{g^2}\Big)\cdot 10^{-2}\,. 
\end{equation}
Since the anticipated error of future lepton colliders\cite{deBlas:2022ofj} measurements of this quantity are estimated to be at the level of $\simeq 1.7\%$, this shift might be observable at the 
$2-3\sigma$ level if the $c_i$ are sufficiently large in comparison to $g^2\simeq 0.43$, but clearly this depends exactly where one ends up in the model parameter space. It should be noted that 
other reasonable choices of $m_{1,2}$ will not significantly alter these basic conclusions.

\section{Summary, Discussion and Conclusions}

The SM faces many theoretical puzzles, \eg, the nature of dark matter's interaction with ordinary matter, as well as an apparently growing number of experimental challenges, 
some fraction of which may be signals for new physics, such as the recent high-precision $W$ mass measurement by CDFII. It behooves us to explore the possibility that some of 
these multiple issues might be inter-related, having solutions which share common elements and this has been one of the objectives of this analysis. In this paper, we have constructed a simple 
toy model that relates the apparent shift in the $W$ mass away from SM expectations to the mass of the dark photon, the mediator responsible for the interaction between us and DM in 
the kinetic mixing portal scenario. As is well-known, the KM portal offers an attractive and testable mechanism to generate a small but phenomenologically interesting coupling between 
the fields of the SM and the dark sector, allowing the DM to reach its observed relic density by thermal means for mass scales of roughly $\lsim 1$ GeV or so. A necessary ingredient of this 
setup is the existence of portal matter fields which simultaneously have both dark and SM couplings that can generate this KM via one-loop diagrams. If the dark gauge group is simply a 
$U(1)_D$ and the PM fields generating this KM are also Higgs-like scalars obtaining the small vevs which also break this $U(1)_D$, then the upper bounds on masses of these fields are 
set by the 
SM isodoublet vev, $\sim 246$ GeV, implying that they should be kinematically accessible at the LHC. From past studies it is known that the generic hallmark signature for PM within 
this class of models is similar in nature to gaugino-like SUSY with observable final states consisting of one or more SM gauge bosons together with MET. 

In the present setup, an $SU(2)_L$, $Y=0$, complex scalar isotriplet and a complex isosinglet, both with $Q_D=1$, simultaneously act as both PM and also obtain these small $\sim1$ GeV 
vevs thus generating the mass of the dark photon as well as a shift in the value of the $W$ boson mass away from the expectations of the SM, matching the measurement from 
CDF II as part of a 
global fit. Clearly, these two effects are correlated in such a framework as has been presented here and KM is then realized in a non-abelian manner between the SM $Z$ and the dark photon 
due to the mass splitting within the PM scalar triplet and can occur at the $\epsilon \gsim 2g_D/e\cdot 10^{-4}$ level. In this setup, the dominant direct experimental signatures for the new 
PM scalars 
were shown to arise in both the $W^\pm+$MET and the $W^+W^-+$ MET channels due to the associated and pair production of the lighter charged scalar, $\Sigma_1$, respectively. These are 
similar in nature to the single and pair production of charged winos and a recast of those searches from ATLAS was shown to provide a lower bound on the the $\Sigma_1$ mass of roughly 
$m_1\geq 230$ GeV. Given the mass relationships between the scalar PM fields within the model, which were shown to essentially depend upon only 3 parameters to a good approximation, the 
perturbativity constraints on the quartic couplings in the scalar potential, the mass bound from ATLAS as well as the constraint on the value of $\epsilon$, a scan of the model parameter space 
was performed. Here, it was shown the mass of the lighter charged PM scalar, $\Sigma_1$ was constrained to lie below roughly $\simeq 385$ GeV while the heavier charged PM 
scalar, $\Sigma_2$, 
as well as the heavy neutral CP-odd and CP-even PM scalars all had their masses constrained to lie below roughly $\simeq 630$ GeV. It is more than likely that the various searches at the 
HL-LHC will be able to cover most if not all of this model's parameter space.

Lastly, we may speculate on how the existence of these new PM scalars may influence the physics of the DM itself. As is well-known, in the traditional KM scenario the DM relic density and its 
direct detection cross section at underground experiments, \eg, are completely controlled by the DM's interaction with the fields of the SM via the exchange of a DP. Naively, at tree-level, 
the existence of PM states with masses far in excess of the typical $\sim1$ GeV physics scale one encounters in such scenarios can have very little direct influence on this physics up to tiny 
mixing effects and this would certainly seem to be the case in the present model as described above. In any setup where the DM is a also complex scalar ($\chi$, which 
doesn't get a vev), one can always write down new quartic interactions of the form $(\lambda_1 \Sigma^\dagger \Sigma+\lambda_2 S^\dagger S+\lambda_3\phi^\dagger \phi)\chi^\dagger \chi$ 
which in the mass eigenstate basis yields the trilinear couplings such as $\sim (h_D,h_{SM}, H)\chi^\dagger \chi$ that can lead to some modifications to this simple interaction picture. 
Of course, apart from the $\lambda_1$ term, such interactions already occur in the familiar dark Higgs singlet model\cite{Wojcik:2021xki} so, as far as this aspect is concerned, there is not 
much that is new in the present setup. To go further we need to speculate on the structure of a (more) 
UV-complete scenario that might involve these PM fields more generally; unfortunately the well-explored set of such models is currently rather limited\cite{Rueter:2019wdf}. However, an 
important common feature of models of this kind is the embedding of the abelian $U(1)_D$ gauge group into a non-abelian structure wherein (some) SM and (some) PM fields may lie in common 
representations so that they may be linked by the new gauge interactions of the non-abelian gauge group. Such a possibility\cite{Rizzo:2021lob}, though certainly not well-explored in full 
detail, may lead to new interactions between the DM and SM fields beyond the familiar ones generated by KM thus possibly altering the familiar results for relic density calculations and direct 
detection experiments in a significant manner. However, further speculation in this direction, as applied to a generalization of the current setup, is beyond the scope of the present work.

Potential experimental anomalies with respect to the predictions of the SM are always valuable as they allow us to explore new ideas and regions of BSM parameter space which we would 
not ordinarily consider in order to examine what the possibilities are that might be realized in nature. Hopefully one of the existing anomalies will lead us to a real discovery in the near future.

\section {Acknowledgements}
The author would like to thank J.L. Hewett for discussions. This work was supported by the Department of Energy, Contract DE-AC02-76SF00515.

\end{document}